\documentstyle[12pt,epsf]{article}

\begin{document}


\def\sst#1{\scriptscriptstyle{#1}}
\def\xf{x_{\!\scriptscriptstyle{f}}}
\def\thf{\theta_{\!\scriptscriptstyle{f}}}
\def\sla#1{\mbox{$#1\hspace*{-0.17cm}\scriptstyle{/}\:$}}
\newcommand{\nc}{\newcommand}
\nc{\postscript}[2] 
{\setlength{\epsfxsize}{#2\hsize}\centerline{\epsfbox{#1}}}
\nc{\bg}{B. Grzadkowski}
\nc{\non}{\nonumber}
\nc{\barx}{\bar{x}}\nc{\pbarn}{\;\hbox {pb}}\nc{\fbarn}{\;\hbox {fb}}
\nc{\hc}{\hbox {h.c.}} 
\nc{\re}{\hbox {Re}} 
\nc{\im}{\hbox {Im}}
\nc{\mev}{\hbox {MeV}} \nc{\gev}{\;\hbox {GeV}} 
\nc{\tev}{\;\hbox {TeV}}
\def\gesim{\lower0.5ex\hbox{$\:\buildrel >\over\sim\:$}} 
\def\lesim{\lower0.5ex\hbox{$\:\buildrel <\over\sim\:$}} 
\nc{\prd}[3]{{\it Phys.\ Rev.}\ {{\bf D{#1}} (#2), #3}}
\nc{\prl}[3]{{\it Phys.\ Rev.\ Lett.}\ {{\bf {#1}} (#2), #3}}
\nc{\plb}[3]{{\it Phys.\ Lett.}\ {{\bf B{#1}} (#2), #3}}
\nc{\npb}[3]{{\it Nucl.\ Phys.}\ {{\bf B{#1}} (#2), #3}}
\nc{\ptp}[3]{{\it Prog.\ Theor.\ Phys.}\ {{\bf {#1}} (#2), #3}}
\nc{\zfp}[3]{{\it Z.\ Phys.}\ {{\bf C{#1}} (#2), #3}}
\nc{\mpla}[3]{{\it Mod.\ Phys.\ Lett.}\ {{\bf A{#1}} (#2), #3}}
\nc{\rmp}[3]{{\it Rev.\ Mod.\ Phys.}\ {{\bf {#1}} (#2), #3}}
\nc{\ijmpa}[3]{{\it Int.\ J.\ Mod.\ Phys.}\
               {{\bf A{#1}} (#2), #3}}
\nc{\ttbar}{t\bar{t}}         \nc{\bbbar}{b\bar{b}}
\nc{\tanb}{\tan \beta}        \nc{\twbdec}{t\to W^+ b}
\nc{\tbwbdec}{\bar{t}\to W^- \bar{b}}
\nc{\epem}{e^+e^-}            \nc{\eett}{\epem \to \ttbar}
\nc{\sigeett}{\sigma_{e\bar{e}\to\ttbar}}
\nc{\wpwm}{W^+W^-}            \nc{\tbar}{\bar{t}}
\nc{\bbar}{\bar{b}}           \nc{\wpp}{W^+}
\nc{\mt}{m_t}    \nc{\mts}{m_t^2}   \nc{\mw}{M_W}    \nc{\mws}{M_W^2}
\nc{\mz}{M_Z}    \nc{\mzs}{M_Z^2}
\nc{\ttbardec}{\ttbar \to W^+W^-\bbbar}
\nc{\wwbb}{W^+W^-\bbbar}      \nc{\sm}{SM}
\nc{\cw}{\cos\theta_W}        \nc{\sw}{\sin\theta_W}
\nc{\sws}{\sin^2\theta_W}     \nc{\sig}{\sigma_{tot}}
\nc{\lp}{{\ell}^+}              
\nc{\lm}{{\ell}^-}
\nc{\lpm}{{\ell}^\pm}
\nc{\tb}{\stackrel{{\scriptscriptstyle (-)}}{t}}
\nc{\bb}{\stackrel{{\scriptscriptstyle (-)}}{b}}
\nc{\fb}{\stackrel{{\scriptscriptstyle (-)}}{f}}
\nc{\epsl}{\epsilon_L}        \nc{\cp}{C\!P}

\nc{\splus}{s_+}       \nc{\smin}{s_-}        \nc{\eps}{\epsilon}
\nc{\psp}{Ps_+}        \nc{\psm}{Ps_-}        \nc{\lsp}{ls_+}
\nc{\lsm}{ls_-}        \nc{\sss}{s_+s_-}      \nc{\m}{m_t}
\nc{\mq}{m_t^2}        \nc{\mr}{\frac{1}{\m}} \nc{\av}{A_{\gamma}}
\nc{\bv}{B_{\gamma}}   \nc{\az}{A_Z}          \nc{\bz}{B_Z}
\nc{\avs}{A_{\gamma}^2}\nc{\azs}{A_Z^2}       \nc{\bzs}{B_Z^2}
\nc{\dav}{\delta \! A_{\gamma}}   \nc{\dbv}{\delta \! B_{\gamma}}
\nc{\dcv}{\delta C_{\gamma}}      \nc{\ddv}{\delta \! D_{\gamma}}
\nc{\daz}{\delta \! A_Z}          \nc{\dbz}{\delta \! B_Z}
\nc{\dcz}{\delta C_Z}             \nc{\ddz}{\delta \! D_Z}
\nc{\dev}{\delta \! E_{\gamma}}   \nc{\dez}{\delta \! E_Z}
\nc{\dfv}{\delta \! F_{\gamma}}   \nc{\dfz}{\delta \! F_Z}
\nc{\rdav}{{\rm Re}(\delta \! A_{\gamma}) \:}
\nc{\rdbv}{{\rm Re}(\delta \! B_{\gamma}) \:}
\nc{\rdcv}{{\rm Re}(\delta C_{\gamma}) \:}
\nc{\rddv}{{\rm Re}(\delta \! D_{\gamma}) \:}
\nc{\rdaz}{{\rm Re}(\delta \! A_Z) \:}
\nc{\rdbz}{{\rm Re}(\delta \! B_Z) \:}
\nc{\rdcz}{{\rm Re}(\delta C_Z) \:}
\nc{\rddz}{{\rm Re}(\delta \! D_Z) \:}
\nc{\idav}{{\rm Im}(\delta \! A_{\gamma}) \:}
\nc{\idbv}{{\rm Im}(\delta \! B_{\gamma}) \:}
\nc{\idcv}{{\rm Im}(\delta C_{\gamma}) \:}
\nc{\iddv}{{\rm Im}(\delta \! D_{\gamma}) \:}
\nc{\idaz}{{\rm Im}(\delta \! A_Z) \:}
\nc{\idbz}{{\rm Im}(\delta \! B_Z) \:}
\nc{\idcz}{{\rm Im}(\delta C_Z) \:}
\nc{\iddz}{{\rm Im}(\delta \! D_Z) \:}
\nc{\cz}{(1+v_e^2)d\:\!'^2}         \nc{\ci}{v_ed\:\!'}
\nc{\ccz}{v_ed\:\!'^2}              \nc{\cci}{d\:\!'}
\nc{\dxdcos}{{d^2\sigma \over d\xf\;d\cos\thf}}
\nc{\gl}{{\mit\Gamma}_{\ell}}         \nc{\gw}{{\mit\Gamma}_W}
\nc{\gf}{{\mit\Gamma}_{\sst{f}}}      \nc{\gb}{{\mit\Gamma}_b} 
\nc{\reaf}{\re(f_2^R)}
\nc{\bet}{\beta}                \nc{\bs}{\bet^2}

\nc{\lspace}{\;\;\;\;\;\;\;\;\;\;}  \nc{\llspace}{\lspace \lspace}

\nc{\beq}{\begin{equation}}   \nc{\eeq}{\end{equation}}
\nc{\bear}{\begin{eqnarray}}   \nc{\eear}{\end{eqnarray}}
\nc{\baa}{\begin{array}}      \nc{\eaa}{\end{array}}
\nc{\bit}{\begin{itemize}}    \nc{\eit}{\end{itemize}}
\nc{\ben}{\begin{enumerate}}  \nc{\een}{\end{enumerate}}
\nc{\bce}{\begin{center}}     \nc{\ece}{\end{center}}

\def\mib#1{\mbox{\boldmath $#1$}}
\def\bra#1{\langle #1 |}      \def\ket#1{|#1\rangle}
\def\vev#1{\langle #1\rangle} \def\dps{\displaystyle}


\vspace*{-1.5cm}
\begin{flushright}
$\vcenter{
\hbox{IFT-11-00}
\hbox{TOKUSHIMA Report} 
\hbox{March, 2000}
}$
\end{flushright}

\vskip 0.5cm
\begin{center}
{\large\bf Signals of CP Violation in Distributions of {\Large $\ttbar$} Decay 
Products at Linear Colliders}
\end{center}

\vspace*{1cm}
\begin{center}
\renewcommand{\thefootnote}{\alph{footnote})}
{\sc Bohdan GRZADKOWSKI$^{\:1),\:}$}\footnote{E-mail address:
\tt bohdan.grzadkowski@fuw.edu.pl}\ and\ 
{\sc Zenr\=o HIOKI$^{\:2),\:}$}\footnote{E-mail address:
\tt hioki@ias.tokushima-u.ac.jp}
\end{center}

\vspace*{1cm}
\centerline{\sl $1)$ Institute of Theoretical Physics,\ Warsaw 
University}
\centerline{\sl Ho\.za 69, PL-00-681 Warsaw, POLAND}

\vskip 0.3cm
\centerline{\sl $2)$ Institute of Theoretical Physics,\ 
University of Tokushima}
\centerline{\sl Tokushima 770-8502, JAPAN}
\vspace*{1cm}

\renewcommand{\thefootnote}{\sharp\arabic{footnote}}
\setcounter{footnote}{0}

\centerline{ABSTRACT}
Angular and energy distributions for leptons and bottom quarks in the
process $\epem \to \ttbar \to {\ell}^\pm/\!\bb \cdots$ have been
calculated assuming the most general top-quark couplings. The double
distributions depend both on modification of the $\ttbar$ production
and $\tb \to \bb\!W$ decay vertices. However, the leptonic angular
distribution turned out to be insensitive to non-standard
parts of $Wtb$ vertex. The method of optimal observables have been
used to estimate sensitivity of future measurements at linear $\epem$ colliders
to top-quark couplings.\\

\vfill


\pagestyle{empty}
\newpage
\pagestyle{plain} 
\setcounter{footnote}{0}

\section{Introduction}

In spite of the fact that the top quark has been discovered already 
several years ago\cite{top} its interactions are still unknown. Therefore it
remains an open question if the top-quark couplings obey the Standard
Model (SM) scheme of the electroweak forces or there exists a
contribution from physics beyond the SM. In this talk\footnote{Presented by B.~Grzadkowski}
I will try to use angular 
and energy distributions of top-quark decay products in the process  
$\epem \to \ttbar \to {\ell}^\pm/\!\bb \cdots$ in order to estimate
how precisely top quark couplings could be determined at future linear collider.

We will parameterize $\ttbar$ couplings to the photon and the $Z$
boson in the following way
\begin{equation}
{\mit\Gamma}_{vt\bar{t}}^{\mu}=
\frac{g}{2}\,\bar{u}(p_t)\,\Bigl[\,\gamma^\mu \{A_v+\delta\!A_v
-(B_v+\delta\!B_v) \gamma_5 \}
+\frac{(p_t-p_{\bar{t}})^\mu}{2m_t}(\delta C_v-\delta\!D_v\gamma_5)
\,\Bigr]\,v(p_{\bar{t}}),\ \label{ff}
\label{prodff}
\end{equation}
where $g$ denotes the $SU(2)$ gauge coupling constant, $v=\gamma,Z$,
and 
\[
\av=\frac43\sw,\ \ \bv=0,\ \ 
\az=\frac1{2\cw}\Bigl(1-\frac83\sin^2\theta_W\Bigr),\ \ 
\bz=\frac1{2\cw}
\]
denote the SM contributions to the vertices. Among the above non-SM
form factors, $\delta\!A_{\gamma,Z}$, $\delta\!B_{\gamma,Z}$, $\delta
C_{\gamma,Z}$ describe $C\!P$-conserving while $\delta\!D_{\gamma,Z}$
parameterizes $C\!P$-violating interactions. 

Similarly, we will adopt
the following parameterization of the $Wtb$ vertex suitable for the
$t$ and $\tbar$ decays:
\begin{eqnarray}
&&\!\!{\mit\Gamma}^{\mu}_{Wtb}=-{g\over\sqrt{2}}V_{tb}\:
\bar{u}(p_b)\biggl[\,\gamma^{\mu}(f_1^L P_L +f_1^R P_R)
-{{i\sigma^{\mu\nu}k_{\nu}}\over M_W}
(f_2^L P_L +f_2^R P_R)\,\biggr]u(p_t), \non \\
\label{dectff}
&&\!\!\bar{\mit\Gamma}^{\mu}_{Wtb}=-{g\over\sqrt{2}}V_{tb}^*\:
\bar{v}(p_{\bar{t}})
\biggl[\,\gamma^{\mu}(\bar{f}_1^L P_L +\bar{f}_1^R P_R)
-{{i\sigma^{\mu\nu}k_{\nu}}\over M_W}
(\bar{f}_2^L P_L +\bar{f}_2^R P_R)\,\biggr]v(p_{\bar{b}}),\non \\
\label{dectbarff}
\end{eqnarray}
where $P_{L/R}=(1\mp\gamma_5)/2$, $V_{tb}$ is the $(tb)$ element of
the Kobayashi-Maskawa matrix and $k$ is the momentum of $W$.

It will be assumed here that interactions of leptons with
gauge bosons are properly described by the SM. Through the
calculations all fermions except the top quark will be considered as
massless. We will also neglect terms quadratic in non-standard form
factors.

\section{Angular and Energy Distributions}

In this section we will present results for $d^2\sigma/d\xf d\cos
\thf$ of the top-quark decay product $f$, where $f$ could be either
$\lpm$ or $\bb$, $\xf$ denotes the normalized energy of $f$ and
$\thf$ is the angle between the $e^-$ beam direction and the
direction of $f$ momentum in the $\epem$ CM frame. 

Using the technique developed by Kawasaki, Shirafuji and Tsai\cite{ts} 
and adopting the general
formula for the $t\bar{t}$ distribution $d\sigma(s_+,s_-)/d{\mit
\Omega}_t$ found by Brzezinski {\it et al.}\cite{BGH}, one obtains the following result for
the distribution:
\beq
\frac{d^2\sigma}{d\xf d\cos\thf}
=\frac{3\pi\beta\alpha_{\mbox{\tiny EM}}^2}{2s}B_{\sst{f}}\:
\Bigl[\:{\mit\Theta}_0^{\sst{f}}(\xf)
+\cos\thf\,{\mit\Theta}_1^{\sst{f}}(\xf)
+\cos^2\thf\,{\mit\Theta}_2^{\sst{f}}(\xf) \:\Bigr],
\label{dis1}
\eeq
where $\beta$ is the top velocity, $\alpha_{\mbox{\tiny EM}}$ is the
fine structure constant and $B_{\sst{f}}$ denotes the appropriate
branching fraction. The energy dependence is specified by the
functions ${\mit\Theta}_i^{\sst{f}}(\xf)$, explicit forms of which
could be found in a recent paper by the authors of this contribution\cite{GH_99}. 
They are parameterized both by production and
decay form factors.

The angular distribution $d\sigma/d\cos\thf$ 
for $f$ could be easy obtained from eq.(\ref{dis1}) 
by the integration over the energy of $f$. {\it It turns out that for $f=\ell$ all 
the non-standard contributions from the decay vertex disappear upon 
integration over the energy $\xf$.}
\begin{figure}[t]
\postscript{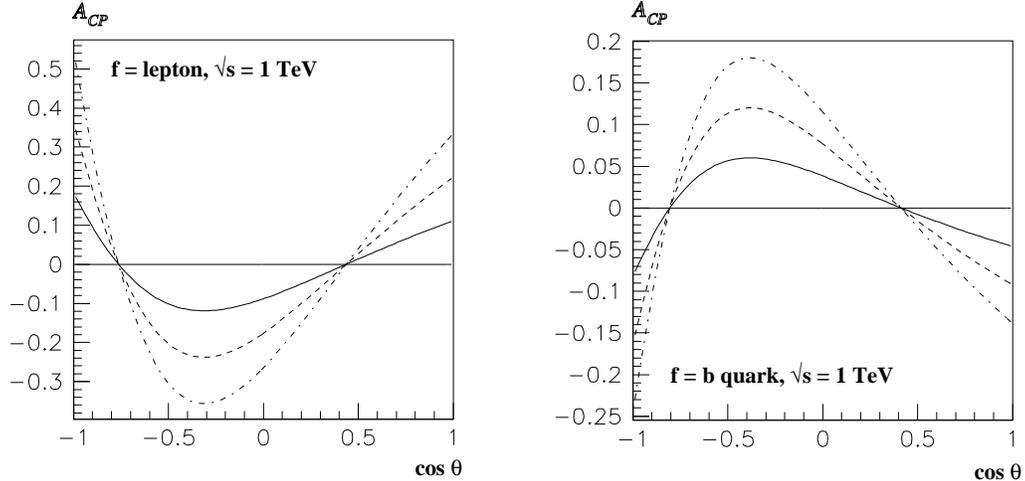}{1.}
\vspace{-0.5cm}
\caption{The $C\!P$-violating asymmetry ${\cal A}_{\sst{CP}}(\thf)$
defined in eq.(\protect\ref{asym}) as a function of $\cos\thf$ for
leptonic and $b$-quark distributions for ${\rm Re}(\delta\!D_\gamma)=
{\rm Re}(\delta\!D_Z)={\rm Re}(f_2^R-\bar{f}_2^L)=$0.1 (solid line),
0.2 (dashed line), 0.3 (dash-dotted line) at $\protect\sqrt{s}=1\tev$
collider energy.  \label{cpasym}}
\vspace{-5mm}
\end{figure}
The fact that the angular leptonic distribution is insensitive to
corrections to the $V$-$A$ structure of the decay vertex allows for
much more clear tests of the production vertices through a measurement
of the distribution, since that way we can avoid a contamination from
non-standard structure of the decay vertex. As an illustration, we
define a $C\!P$-violating asymmetry which could be constructed using
the angular distributions of $f$ and $\bar{f}$:
\beq
{\cal A}_{\sst{CP}}(\thf)= \Big[\:
{\displaystyle \frac{d\sigma^+(\thf)}{d\cos\thf}-
\frac{d\sigma^-(\pi-\thf)}{d\cos\thf}}
\:\Bigr]\Big/\Bigl[\:
{\displaystyle \frac{d\sigma^+(\thf)}{d\cos\thf}+
\frac{d\sigma^-(\pi-\thf)}{d\cos\thf}}
\:\Bigr], \label{asym}
\eeq
where $d\sigma^{+/-}$ is referring to $f$ and $\bar{f}$
distributions, respectively.
The asymmetry for $f={\ell}$ is
sensitive to $C\!P$ violation originating exclusively from the
production mechanism, i.e. $C\!P$-violating form factors $\delta\!D_\gamma$
and $\delta\!D_Z$ while the decay vertex enters with the SM
$C\!P$-conserving coupling. For bottom quarks the effect of the
modification of the decay vertex is contained in corrections to so called 
depolarization factors\cite{GH_99}, $\alpha^b+\alpha^{\bar{b}}\sim
{\rm Re}(f_2^R-\bar{f}_2^L)$ with SM $C\!P$-conserving
contribution from the production process.
As it is seen from fig.\ref{cpasym} for $\sqrt{s}=1\tev$ the
asymmetry could be quite large, e.g., reaching for the semileptonic
decays $\sim 20\%$ for ${\rm Re}(\delta\!D_\gamma)={\rm Re}(\delta\!
D_Z)=0.2$.

\section{Optimal Observable Analysis}

Using the double angular and energy distributions
an expected statistical uncertainty for determination of real parts 
for all the form factors has been found
adopting optimal observables\cite{oo} and varying the
beam polarizations $P_{e^-}$ and $P_{e^+}$. $|\cos\theta|\leq 0.9$
has been assumed as a cut for the polar angle. For $\ttbar$ tagging 
efficiency in $\ell$ + 4 jet channel we adopted  60\% and for the integrated luminosity
we chose the TESLA design with $L=500\fbarn^{-1}$ at $\sqrt{s}=500\gev$. 

Generically we have observed 
that positive polarization led to smaller statistical errors for the eight form
factors in the production vertices. For each form factor we have adjusted
the optimal beam polarization such that the statistical error was minimal:
\begin{equation}
\begin{array}{ll}
{\mit\Delta}[\:\re(\delta A_\gamma)\:]=0.16\ \ \ \ \ \ \ &  {\rm for}\ P_{e^-}=0.7\ {\rm and}\ P_{e^+}=0.7 \\
{\mit\Delta}[\:\re(\delta A_Z)\:]=0.07 & {\rm for}\ P_{e^-}=0.5\ {\rm and}\ P_{e^+}=0.4 \\
{\mit\Delta}[\:\re(\delta B_\gamma)\:]=0.09\ \ \ \ \ \ \ &
  {\rm for}\ P_{e^-}=0.2\ {\rm and}\ P_{e^+}=0.2 \\
{\mit\Delta}[\:\re(\delta B_Z)\:]=0.27 &
  {\rm for}\ P_{e^-}=0.4\ {\rm and}\ P_{e^+}=0.4 \\
{\mit\Delta}[\:\re(\delta C_\gamma)\:]=0.11\ \ \ \ \ \ \ &
  {\rm for}\ P_{e^-}=0.1\ {\rm and}\ P_{e^+}=0.0 \\
{\mit\Delta}[\:\re(\delta C_Z)\:]=1.11 &
  {\rm for}\ P_{e^-}=0.1\ {\rm and}\ P_{e^+}=0.0 \\
{\mit\Delta}[\:\re(\delta D_\gamma)\:]=0.08\ \ \ \ \ \ \ &
  {\rm for}\ P_{e^-}=0.2\ {\rm and}\ P_{e^+}=0.1 \\
{\mit\Delta}[\:\re(\delta D_Z)\:]=14.4 &
  {\rm for}\ P_{e^-}=0.2\ {\rm and}\ P_{e^+}=0.1 \\
\end{array}
\end{equation}

As it is seen the precision of $\delta\{C,D\}_Z$ measurement would be 
very poor even for the optimal 
polarization. In addition, determination of $\delta D_\gamma$ would be difficult
as well since its error varies rapidly with the polarization.
For example, ${\mit\Delta}[\:\re(\delta D_\gamma)\:]$ becomes
0.86 for $P_{e^-}=0.1/P_{e^+}=0.1$ and 0.99 for $P_{e^-}=0.3/P_{e^+}=0.1$.
The source of that sensitivity is hidden in the neutral current structure 
with $\sin^2\theta_W\simeq 0.23$. Indeed, the optimal polarization 
becomes $P_{e^-}=0.1$ instead of 0.2 $({\mit\Delta}[\:\re(\delta D_\gamma)\:]=0.09)$ 
for $\sin^2\theta_W=0.25$.
On the other hand, a good determination (almost independently of the polarization) 
could be expected for $f_2^R$. Indeed, the best precision is
\begin{equation}
{\mit\Delta}[\:\re(f_2^R)\:]=0.01\ \ \ \ \ \ \ 
  {\rm for}\ P_{e^-}=-0.8\ {\rm and}\ P_{e^+}=-0.8
\end{equation}
however, we have ${\mit\Delta}[\:\re(f_2^R)\:]=0.03$ even for $P_{e^-}=
P_{e^+}=0$.

\section{Summary and Conclusions}
We have presented here the angular and energy distributions for
$\fb$ in the process $\epem \to \ttbar \to \fb \cdots$, where $f=
\ell$ or $b$ quark. The most general ($C\!P$-violating {\it
and} $C\!P$-conserving) couplings for $\gamma\ttbar$, $Z \ttbar$ and
$Wtb$ have been assumed.
The bottom-quark mass has been neglected and we have kept only
terms linear in anomalous couplings.

Test of CP violation has also been discussed, introducing a CP-sensitive asymmetry
${\cal A}_{CP}$ as an example.

Using the double angular and energy distribution of a lepton we have found that
at $\sqrt{s}=500\gev$ with the integrated luminosity $L=500\fbarn^{-1}$
the best determined top-quark coupling would be the axial coupling of the Z boson 
with the error ${\mit\Delta}[\:\re(\delta A_Z)\:]=0.07$  while the lowest
precision is expected for $\re(\delta D_Z)$ with ${\mit\Delta}[\:\re(\delta D_Z)\:]=14.4$.

\section*{Note added}
After this work has been presented, a paper by Rindani\cite{rindani} appeared where
the double angular and energy leptonic distribution has been also found.

\section*{Acknowledgments}
BG is grateful to the organizers of the PASCOS99 conference
for creating a very warm and inspiring atmosphere during the meeting.
This work is supported in part by the State
Committee for Scientific Research (Poland) under grant 2~P03B~014~14
and by Maria Sk\l odowska-Curie Joint Fund II (Poland-USA) under
grant MEN/NSF-96-252.

\end{document}